\documentclass[12pt]{iopart}
\usepackage{graphicx}
\usepackage{hyperref}
\hypersetup{hidelinks}

\usepackage{iopams}
\expandafter\let\csname equation*\endcsname\relax
\expandafter\let\csname endequation*\endcsname\relax
\usepackage{amsmath}

\usepackage{multirow}
\usepackage{amssymb}
\usepackage{bbm}
\usepackage{color}
\usepackage{dcolumn}
\usepackage{bm}
\usepackage{float}
\usepackage[normalem]{ulem}
\usepackage{braket}
\usepackage{changepage}
\usepackage[subrefformat=parens,labelformat=parens,caption=false]{subfig}
\usepackage{bbm}
\usepackage[dvipsnames]{xcolor}
\usepackage{esint}
\usepackage{lineno}
\usepackage{verbatim}
\usepackage{bm}
\usepackage{bbold}
\usepackage{framed}

 \usepackage[inner=2.7cm, outer=2cm, bottom=2cm, top=3cm]{geometry}

\AtBeginEnvironment{figure}{\mathindent=0pt }

\newcommand\norm[1]{\left\lVert#1\right\rVert}

\begin{document}

\title{Performance and scaling analysis of variational quantum simulation}

\author{Mario Ponce\textsuperscript{\normalfont{1,2}}, Thomas Cope\textsuperscript{\normalfont{1}}, In\'es de Vega\textsuperscript{\normalfont{1,2}}, Martin Leib\textsuperscript{\normalfont{1}}}
\address{\textsuperscript{1} IQM Quantum Computers, Georg-Brauchle-Ring 23-25, 80992 Munich, Germany}
\address{\textsuperscript{2} Department of Physics and Arnold Sommerfeld Center for Theoretical Physics, Ludwig-Maximilians-Universit\" at M\" unchen, Theresienstrasse 37, 80333 Munich, Germany}
\eads{\mailto{mario.ponce@meetiqm.com}}
\begin{abstract}
We present an empirical analysis of the scaling of the minimal quantum circuit depth required for a variational quantum simulation (VQS) method to obtain a solution to the time evolution of a quantum system within a predefined error tolerance. In a comparison against a non-variational method based on Trotterized time evolution, we observe a better scaling of the depth requirements using the VQS approach with respect to both the size of the system and the simulated time. Results are also put into perspective by discussing the corresponding classical complexity required for VQS. Our results allow us to identify a possible advantage region for VQS over Trotterization.
\end{abstract}
\keywords{variational quantum algorithms, time evolution, scaling analysis}

\section{Introduction}

In recent years, quantum computing has emerged as a promising avenue for tackling complex computational tasks surpassing the capabilities of classical computers~\cite{preskill}. Among the many possible applications of this new technology, simulating the time evolution of quantum systems stands as a cornerstone problem with profound implications across disciplines; from condensed matter physics to chemistry and beyond~\cite{chemist}. The ability to accurately simulate the dynamics of quantum systems opens doors to understanding fundamental phenomena and, among others,  designing novel materials and molecules with tailored properties.

Traditional computational methods for simulating real-time evolution face significant challenges when applied to large-scale quantum systems due to the exponential growth of the computational resources required~\cite{classical}. In contrast, quantum computers offer a potential solution by harnessing the principles of quantum mechanics to directly simulate quantum systems.

One of the most well-established approaches to quantum time evolution are product formulas~\cite{univ_lloyd, future_ibm,trotter_error,berry_exp}, often referred to as Trotterization. The idea behind this technique is to approximately split the time-evolution operator into smaller unitary operators representable by elementary gates which can be performed on a quantum computer. The approximation may be improved by repeating the process in several layers to account for the non-commutativity of the Hamiltonian. Product formulas are useful because of their intuitive construction as well as the possibility to derive asymptotic complexity expressions due to their well-defined mathematical structure~\cite{asympt}. In general, however, they are not NISQ-friendly, since the circuit depths involved in achieving a specific precision are usually limited by the $T_1$ times of the qubits, that is, by the time that they are able to reliably hold quantum information. As a direct consequence of this, it is generally not feasible to aim for long-time simulations using a product formula in the NISQ era.

This motivates the search for alternative, potentially noise-resilient quantum algorithms for quantum time evolution, such as variational quantum algorithms~\cite{variational}. These algorithms leverage the expressiveness of parametrized quantum circuits to represent quantum states. By iteratively adjusting the parameters, they seek to approximate the dynamical evolution of the target quantum system. In order to accomplish this, a variational principle has to be devised that is capable of approximately transferring the rules for the dynamical evolution of the quantum state to that of the parameters.

Particularly promising for time evolution is the Variational Quantum Simulation (VQS) algorithm~\cite{vqs1}, a hybrid quantum-classical algorithm characterized by an algebraic update rule for the parameters. Its interest is due to the availability of an error bound and the alleged shallowness of the quantum circuits involved. Since its inception in 2017, it has been applied to different scenarios such as state preparation using the quantum imaginary time evolution trick~\cite{vqs2} and open quantum systems~\cite{vqs4}, other than to the standard unitary time evolution. Adaptive strategies have also been proposed following the ADAPT-VQE~\cite{adaptvqe} framework for both real and imaginary time evolution~\cite{adaptvarrte,adaptvarqite}.

There have been several theoretical efforts explaining the properties of VQS, such as a summary of the theory in~\cite{vqs3} and a study of error bounds~\cite{error_bounds}. However, to the best of our knowledge, an analysis of its performance and scaling is lacking in the literature.

Due to the heuristic nature of most variational algorithms, it is hard or impossible to obtain theoretical performance guarantees. Therefore, in this paper, we provide an empirical study of the scaling of VQS and a direct performance comparison to a second-order product formula~\cite{trotter1,trotter2}, to which we refer as Trotterization or simply Trotter. We do this for a specific family of Hamiltonians and draw conclusions in different paradigmatic cases. 

The way in which we perform this comparison is useful not only to get an intuitive understanding of the performance of the variational method, but also as a first step towards the identification of a quantum advantage scenario, as explained in~\cite{future_ibm, Childs_useful}. Identifying a quantum advantage scenario means finding the minimum quantum resources such that the quantum-assisted computation may outperform the classical one. It would not be surprising that quantum advantage is first found by simulating quantum dynamics, since this is arguably the most natural application of quantum computers. A simple spin system with local interactions such as the one we consider on this paper is also a good candidate because it is likely that it involves less overhead than other more sophisticated systems. Lastly, the choice $t_\text{f} = $ $n_\text{q}$ (where $t_\text{f}$ stands for simulated time and $n_\text{q}$ for the size of the system) in our comparison setup, which will appear later on section~\ref{section:results}, helps us to highlight the system-size dependence in the scaling analysis of quantum simulation algorithms, one of the most relevant aspects for practical applications. 

This article is structured as follows: in section~\ref{section:method} we provide a concise introduction to the variational quantum simulation algorithm and to second-order Trotterization,  we introduce the benchmarking setup in section~\ref{section:bench} and show our computational results in section~\ref{section:results}.

\section{Time Evolution Methods}
\label{section:method}

\subsection{Variational Quantum Simulation}

Here we describe a method, which we refer to as Variational Quantum Simulation (VQS), to approximate the time dynamics of a quantum state generated by a  Hamiltonian~\cite{vqs1,vqs3} according to the Schrödinger equation
\begin{equation}
\frac{d}{d t}|\Psi(t)\rangle=-i H|\Psi(t)\rangle.
\end{equation}
The time-evolved quantum state is approximated by a parametrized ansatz
\begin{equation}
\ket{\psi(\bm{\theta}(t))} = \prod_{i} \mathcal{U}_i(\theta_i)\ket{\psi(\bm{\theta}_0)},
\end{equation}
whose parameters $\bm{\theta}=(\theta_i)_i$ need to be updated in order to track the time evolution of the state. This is accomplished with a time-dependent variational principle or McLachlan's principle
\begin{equation}
\dot{\bm{\theta}}(t)=\operatorname{argmin}_{\bm{\theta}} \norm{\left(\frac{d}{d t}+i H\right)|\psi(\bm{\theta}(t))\rangle},
\end{equation}
which leads to an update rule for the parameters in the form of a linear system of ordinary differential equations (ODEs)
\begin{equation}
   \mathcal{A} \bm{\dot{\theta}} = \mathcal{C},
   \label{eq:lse}
\end{equation}
where the dot over the vector of parameters denotes time derivative.
The matrix $\mathcal{A}$, which contains information about the quantum geometry of the ansatz~\cite{qgeom}, is commonly known as the Quantum Fisher Information (QFI) matrix. Its entries correspond to overlaps of derivatives of the ansatz,
\begin{equation}
   \mathcal{A}_{i,j} = \text{Re}\left(\braket{\partial_i \psi |\partial_j \psi} - \braket{\partial_i \psi | \psi}\braket{\psi | \partial_j \psi}\right).
\end{equation}
The vector $\mathcal{C}$ encodes the action of the Hamiltonian on the ansatz,
\begin{equation}
   \mathcal{C}_{i} = \text{Im}\left(\braket{\partial_i \psi | H | \psi} + \braket{\psi | \partial_i \psi}\braket{\psi | H | \psi}\right).
\end{equation}
In the last two equations the dependence of the ansatz with respect to the parameters has been omitted for better readability. Also, the partial derivatives are written inside the kets and bras to make use of the bra-ket notation, but it should be noted that the notation is being abused as the partial derivatives of the parametrized ansatz are not normalized, so they are not proper quantum states.

This algorithm constitutes an example of a hybrid quantum-classical algorithm. The quantum computer is employed to evaluate the entries of the matrix $\mathcal{A}$ and the vector $\mathcal{C}$, consisting of overlaps of the parametrized ansatz and its partial derivatives, which can be represented by circuits. Although these circuits may be shallow, it is likely that it would be costly to evaluate them on classical computers due to the possibly generated entanglement. The linear system of equations and the system of ODEs are then solved approximately one after the other using numerical methods on a classical computer. With that, the parameters are updated and then fed back into the quantum computer for the next time step, and that constitutes a quantum-classical feedback loop.

The inversion of the matrix $\mathcal{A}$ is not trivial because it is usually ill-conditioned, so a direct inversion is not possible in most cases. To mitigate this, one can, for example, apply some technique such as a direct Tikhonov regularization~\cite{tikhonov} or solve the system indirectly with a least-squares solver. We found that the latter works best in general. In particular, we used NumPy's \textit{lstsq} function~\cite{numpy}. We have found that a least-squares linear equation solver with its default parameters (a cut-off ratio of the machine precision times the size of the matrix is imposed on the small singular values) works best for our purposes.

For the time integration of the system of ODEs, we choose an adaptive classical numerical method that can keep the integration error under control by automatically adapting the time step, namely a Runge-Kutta 45 method~\cite{rk45}. In particular, we again employ SciPy's implementation of this method~\cite{scipy}.

An important feature of VQS is that an a-posteriori error bound (sometimes called McLachlan distance) for the fidelity of the solution is readily available within the workflow of the algorithm,
\begin{equation}
\label{eq:error_bound}
\|(d / d t+i H)|\psi(\bm{\theta}(t))\rangle \|^2 = \sum_{i,j} \mathcal{A}_{i,j}^{R} \dot{\theta}_i \dot{\theta}_j-2 \sum_i \mathcal{C}_i^I \dot{\theta}_i+\left\langle H^2\right\rangle,
\end{equation}
where the superscripts $R$ and $I$ denote real and imaginary parts respectively, and $\left\langle H^2\right\rangle=\braket{\psi(\bm{\theta}(t))|H^2|\psi(\bm{\theta}(t))}$. This is a very valuable source of information because the error can be estimated in real time and, if not satisfactory, adjustments can be made to the algorithm to try and improve its performance. Such an adaptive process is essential for real-world applications and will be described below in connection to the specific ansatz that we choose.

\subsection{Trotterization}

For comparison purposes we also study a second-order Trotter-Suzuki product formula~\cite{trotter1,trotter2}. Trotterization is a way to represent the time evolution operator on a quantum computer, which for the evolution dictated by the Schrödinger equation corresponds to an exponential. This exponential cannot generally be identified exactly with the unitaries represented by the quantum gates in a gate-based quantum computer.  The idea is then to split this exponential into terms that are representable by gates in a quantum circuit. If the Hamiltonian has non-commuting terms, the splitting is not trivial in general.

In particular, we use the second-order Trotter product formula, analogous to the Strang splitting used in classical numerical methods, which for a Hamiltonian with two non-commuting terms $H^A$ and $H^B$ is given by
\begin{equation}
\mathrm{e}^{\left(-i\left(H^A + H^B\right)t\right)} \approx \left[ \mathrm{e}^{\left(-iH^Bt/(2n)\right)} \mathrm{e}^{\left(-iH^At/n\right)}
\mathrm{e}^{\left(-iH^Bt/(2n)\right)} \right]^{n} + \mathcal{O}\left((t/n)^3\right),
\end{equation}
where $t$ stands for time, $n$ the number of steps in which the time interval is split, and $\mathcal{O}\left((t/n)^3\right)$ represents terms of third order and higher in the time step $t/n$. The error of this approximation scales as $\mathcal{O}\left((t/n)^3\right)$~\cite{trotter_error}. However, it is important to note that Trotterization errors accumulate over long simulation times, unless they are reduced by decreasing the time step, with a consequent increase in circuit depth. It is therefore clear that the circuit depths involved in Trotterization will monotonically increase with simulated time.

 The second-order Trotter product formula provides arguably the best balance between accuracy and computational cost. While it is more accurate than the first-order formula, it still has limitations in accurately approximating the time evolution operator when $t$ is large or when the norm of the commutator $[H^A, H^B]$ is significant compared to the norms of $H^A$ and $H^B$ individually. Higher-order Trotter product formulas exist to improve accuracy by recursively using lower-order formulas, but they generally come with increased quantum computational cost. Product formulas are also applicable when the Hamiltonian is composed of more than two non-commuting terms.

\section{Setup for Numerical Simulation}
\label{section:bench}

We test the VQS algorithm by means of a noiseless statevector simulation on a nearest-neighbours 1D transverse-field Ising model
\begin{equation}
H=\sum_k a_k X_k + \sum_{\langle i, j\rangle} b_{i,j} Z_i Z_j = H^{A} + H^{B},
\label{eq:ising}
\end{equation}
where 50 instances of the coefficients $a_k$ and $b_{i,j}$ are chosen randomly from the interval $(-1,1)$, and we have defined the two non-commuting blocks of one- and two-body terms as $H^A$ and $H^B$ respectively.

We consider a layered problem-specific ansatz, for which the parametrized unitaries of each layer correspond to the Hamiltonian Pauli strings in~(\ref{eq:ising}),
\begin{equation}
\ket{\psi(\bm{\theta}(t))} \approx \prod \left[ \prod_{m}\exp{\left(-i\theta_m  H^A_{m}\right)} \prod_{n} \exp{\left(-i\theta_n H^B_{n}\right)} \right] \ket{\psi(\bm{\theta_0})},
\end{equation}
where $H^A_m$ and $H^B_n$ denote the individual summands of $H^A$ and $H^B$ respectively, and the first product corresponds to the number of layers.

This type of ansatz is sometimes called a Hamiltonian Variational Ansatz (HVA). Similar ansätze have been used in~\cite{hva1,hva2,hva3}, characterized by having a limited level of expressivity and generality, but being able to offer a good performance for the dynamical subspace induced by the Hamiltonian at hand. We chose it in order to make it easier for VQS to find good solutions for the problem at hand, but also because it has the same structure as a product formula, where the increase in the number of repetitions of the unitaries addresses the non-commutativity of the Hamiltonian.

\begin{figure}[t]
\begin{framed}
\centering
\includegraphics[width=0.6\linewidth]{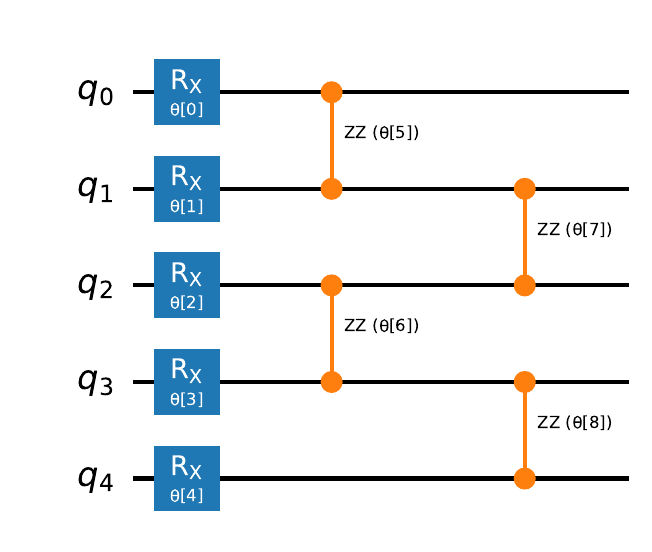}
\caption{First layer of the circuit corresponding to the problem-specific~(\ref{eq:ising}) ansatz that we use for our analysis for 5 qubits. More layers can be added for better results, increasing the total depth and the total number of parameters.}
\label{fig:trotter_ansatz}
\end{framed}
\end{figure}

A representation of the first layer of the ansatz for our Hamiltonian (\ref{eq:ising}) can be seen in figure~\ref{fig:trotter_ansatz}. We chose to distribute the two-qubit parametrized rotations in a brickwall structure, so that the depth of each layer would remain constant instead of growing linearly with the number of qubits, as is the case in the naive cascaded structure. 
We base all our computational experiments on a classical simulation of the quantum-classical feedback loop involved in this algorithm. The quantum part involves the calculation of the entries of the matrix $\mathcal{A}$ and the vector $\mathcal{C}$. These are estimated by the evaluation of overlaps using Qulacs, a fast quantum circuit simulator~\cite{qulacs}. The overlaps are of partial derivatives of the ansatz
\begin{equation}
   \ket{\partial_k \psi(\bm{\theta}(t)) } = \frac{1}{2} \prod_{i} \dots \, \mathcal{U}_{k-1}(\theta_{k-1}) \, \mathcal{U}_{k}(\theta_k + \pi) \, \mathcal{U}_{k+1}(\theta_{k+1}) \dots \, \ket{\psi(\bm{\theta_0})},
\end{equation}
where we have again abused the ket notation as this is not a normalized quantum state.

\section{Results}
\label{section:results}

To understand how VQS performs, we conduct a computational study of its scaling with respect to two important quantities: the number of qubits ($n_\text{q}$) and the simulated time ($t_\text{f}$). As previously stated, for comparison purposes we carry out the same study for a second-order product-formula, to which we refer as Trotter on the different plots.

We choose quantum circuit depth as the main performance metric. This is motivated by the fact that the information processing capability of most quantum computing platforms, such as that based on superconducting qubits, is mainly limited by coherence times~\cite{future_ibm}. The main restricting factor for a quantum algorithm to run on such a platform is therefore that the circuits involved are not too deep, that is, that they can be executed before the qubits stop being able to reliably hold quantum information.

In this study, the minimum depth is found by implementing an adaptive process in which a maximum tolerable error in the fidelity (0.05) of a statevector simulation is fixed beforehand and the solution of the approximation algorithm is compared against an exact solution obtained by direct classical computation using QuTiP~\cite{qutip1,qutip2}. If the error is bigger than the tolerance, the simulation starts over with one more layer in the parameterized ansatz. We again stress that this adaptive process can also be employed in a real application, where the exact solution is not accessible, because of the possibility of computing an error bound~(\ref{eq:error_bound}) and using that as a reference instead of the exact solution. However, to make our results independent from some specific strategy to include the error bound, we decided to rely on a direct numerical calculation of the time dynamics, which we consider as exact, and consider all following results as an upper bound performance for all possible strategies to set up an adaptive technique based on the error bound.  

As mentioned above, we choose the Hamiltonian coefficients randomly, and define a non-trivial initial state by randomly initializing the parameters of a non-trainable first layer of the ansatz. This way, we generate 50 different dynamical problems, defined by both the Hamiltonian and the initial state. We look at the simulation results from two paradigmatic perspectives to illustrate the qualitative behaviour of the scaling with respect to the two most relevant quantities: the number of qubits, that is, the size of the system; and the simulated time $t_\text{f}$. Our simulations go from 2 to 10 qubits and from 1 to 14 time units as simulated time. We chose these ranges to keep running times reasonably low and because the scaling trends in the figures below already become clear.

\subsection{Scaling in the number of qubits}
\label{subsection:fixed_short}

\begin{figure}[t]
\begin{framed}
\centering
\includegraphics[width=0.7\linewidth]{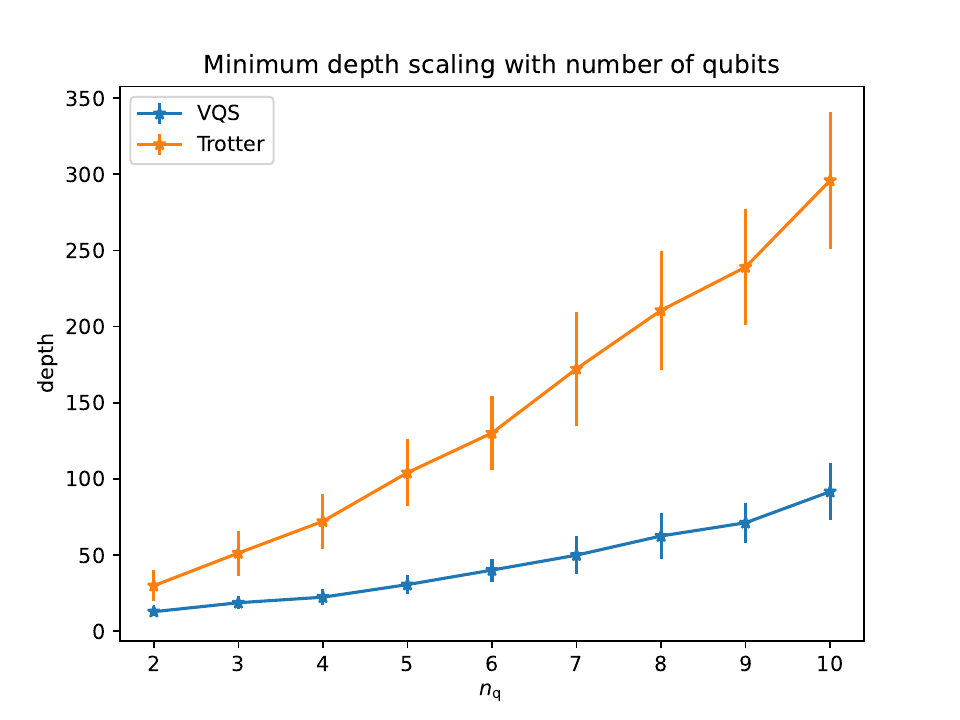}
\caption{Comparison of the minimum depth for the final solution to reach a fidelity above 0.95 for VQS and Trotter for a linearly-increasing simulated time $t_\text{f} = n_\text{q}$. We see that VQS outperforms Trotter, and the trend suggests that it would do so for longer simulated times and larger systems.}
\label{fig:nqubits_comparison}
\end{framed}
\end{figure}

In the first case we set the number of qubits as an independent variable and we take a linearly-increasing simulated time $t_\text{f} = $ $n_\text{q}$ (see figure~\ref{fig:nqubits_comparison}). We observe a lower minimum depth for the variational method in the depicted range, with the trend suggesting that this behavior continues when simulating for longer times. This is reinforced by the fit and extrapolation in figure~\ref{fig:extrapolation_fit} in subsection~\ref{subsection:fitting_extrapolation}, where it is clear that the case highlighted on this subsection falls well within the VQS advantage region.  

\subsection{Scaling in the simulated time}
\label{subsection:long_time}

\begin{figure}[t]
\begin{framed}
\centering
\includegraphics[width=0.7\linewidth]{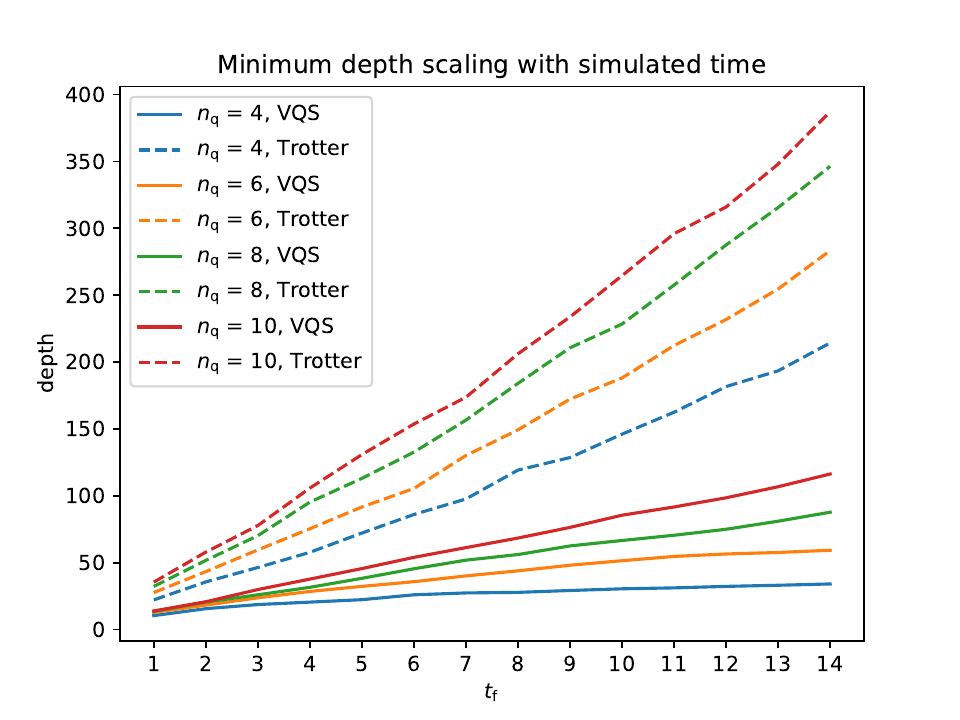}
\caption{Comparison of the minimum depth for the final solution to reach a fidelity above 0.95 for VQS and Trotter for long simulated times. Both algorithms show a growing depth requirement with simulated time, but VQS' growth rate is slower. We present here the results only for even-numbered qubits in the range (3-10) for better visibility, but the qualitative behavior is the same for the odd-numbered cases. This suggests that VQS might be a good alternative to Trotter for long-time quantum simulations.}
\label{fig:tfs_comparison}
\end{framed}
\end{figure}

Secondly, we fix the number of qubits and vary the simulated time, going beyond $t_\text{f}=n_\text{q}$. The results are presented in figure~\ref{fig:tfs_comparison}. We observe how the depth requirement grows steadily for both algorithms as time increases, but VQS shows a tendency for slower depth growth than the corresponding Trotterization. We emphasize that for coherence-limited quantum computing platforms, this better depth performance has the potential to make the variational method feasible where Trotterization is not.

\subsection{Fitting and extrapolation}
\label{subsection:fitting_extrapolation}

\begin{figure}[t]
\begin{framed}
\centering
\includegraphics[width=0.7\linewidth]{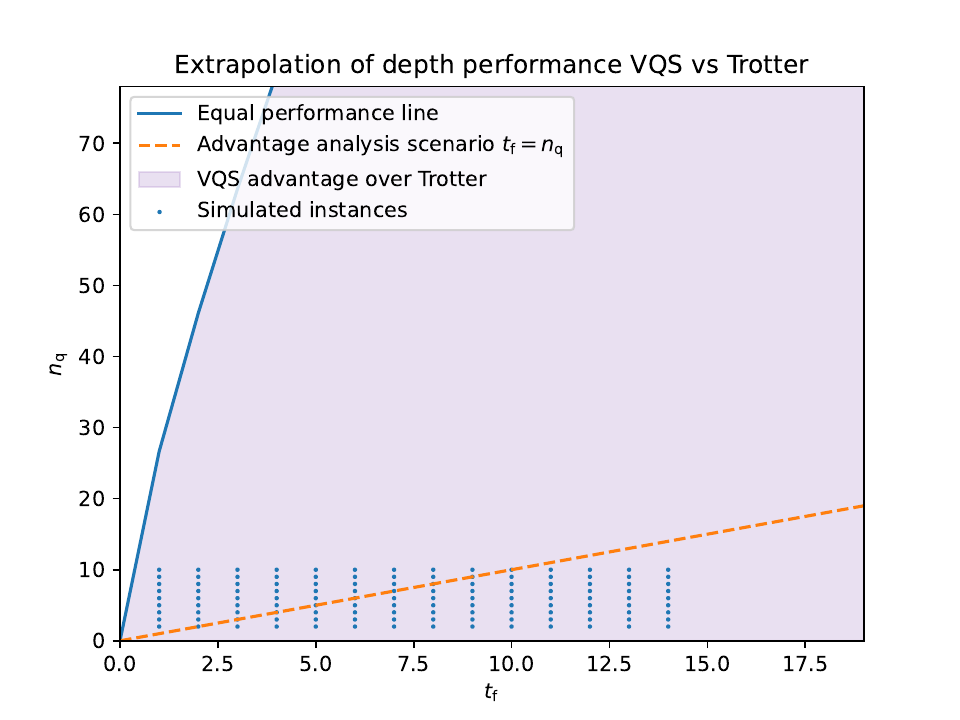}
\caption{Representation of the relative performance of VQS and Trotter in the $t_\text{f}$-$n_\text{q}$ plane. We use a power-law-like function of the form $a \, n_\text{q}^{b} \, t_\text{f}^{c}$ as in~(\ref{fit}) to fit our simulation data and use the resulting function to extrapolate to larger values in the $t_\text{f}$-$n_\text{q}$ plane. The blue line denotes the boundary of equal performance of VQS and Trotter in accordance with the fitted function, the shaded area below is a region where VQS has an advantage over Trotter in terms of depth requirements according to our fitted results. The dots mark the instances that were actually simulated ($n_\text{q}$ in the range 2-10 and $t_\text{f}$ in the range 1-14), some of them appearing directly in figures~\ref{fig:nqubits_comparison} and~\ref{fig:tfs_comparison}. The fit parameters for VQS are $a$ = 1.587 $\pm$ 0.152, $b$ = 0.997 $\pm$ 0.035, $c$ = 0.743 $\pm$ 0.028 and for Trotter $a$ = 3.469 $\pm$ 0.162, $b$ = 0.451 $\pm$ 0.011, $c$ = 1.287 $\pm$ 0.017.}
\label{fig:extrapolation_fit}
\end{framed}
\end{figure}

Upon inspection of the scaling behavior in figures~\ref{fig:nqubits_comparison} and~\ref{fig:tfs_comparison}, from where it seems that the circuit depth scales polynomially with $n_\text{q}$ and $t_\text{f}$, we propose a fitting function of the form 
\begin{equation}
D(n_\text{q},t_\text{f})=a \, n_\text{q}^{b} \, t_\text{f}^{c},
\label{fit}
\end{equation}
where $D$ represents the circuit depth, and a, $b$, and $c$ are the fitting parameters, to all the simulated instances. We find that for VQS these fitting parameters are $a$ = 1.587 $\pm$ 0.152, $b$ = 0.997 $\pm$ 0.035, and $c$ = 0.743 $\pm$ 0.028, while for Trotter they are $a$ = 3.469 $\pm$ 0.162, $b$ = 0.451 $\pm$ 0.011, and $c$ = 1.287 $\pm$ 0.017. We use this fitting to perform an extrapolation and represent this in figure~\ref{fig:extrapolation_fit}. The fitting hints towards a better time scaling for VQS and a better qubit scaling for Trotter. As we mentioned in subsection~\ref{subsection:fixed_short}, the scaling advantage scenario $t_\text{f} = n_\text{q}$ falls well within the region of advantage of VQS over Trotter determined by the fit.

\subsection{Classical computational cost}
\label{subsection:classical_cost}

Since VQS is a hybrid quantum-classical algorithm, after its performance comparison against a purely quantum algorithm based on Trotter, there remains the question of whether the classical cost involved in VQS might be comparable or higher than that of a purely classical algorithm for time evolution. If the hybrid algorithm is well constructed, the classical part of the algorithm should not use more computational resources than a fully classical solution, so that the quantum part contributes positively.

\begin{figure}[t]
\begin{framed}
\centering
\includegraphics[width=0.7\linewidth]{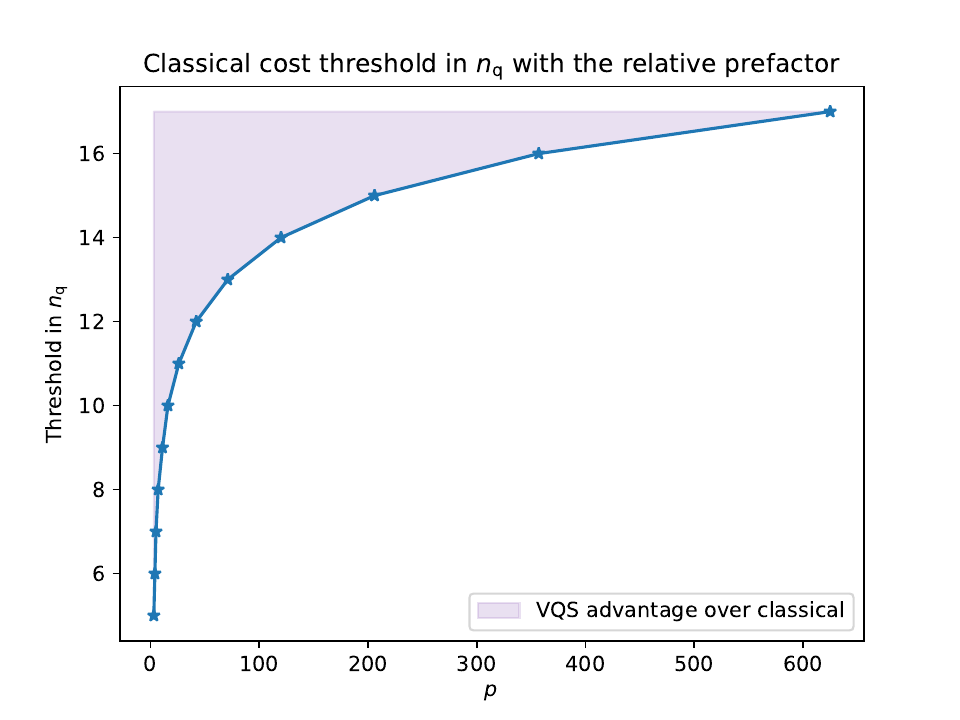}
\caption{Threshold in the number of qubits for which the computational cost of the classical part of VQS times a relative prefactor $p$ is smaller than the computational cost of a purely classical approach based on matrix-vector multiplication (see~(\ref{eq:clcost})), for the case when the simulated time equals the number of qubits.}
\label{fig:threshold}
\end{framed}
\end{figure}

Let us make a simple comparison using the fit from the previous subsection, for the case in which the simulated time equals the number of qubits. We can identify the computational cost of the classical part of VQS with that of inverting a matrix, which is about $m^3$, where $m$ is the number of parameters used in the variational ansatz. Conversely, the cost of a naive classical simulation of time evolution based on matrix-vector multiplications following a Trotter-like structure is asymptotically of the order of $k \, 2^n$, where $k$ is the number of Trotter steps and $n_{\text{q}}$ the number of qubits, making $2^n$ the dimension of the Hilbert space. That is,
\begin{equation}
\label{eq:clcost}
    p \, m^3 < k \, 2^{n_{\text{q}}},
\end{equation}
where we have introduced a relative prefactor $p$, which accounts for some instance-dependent, asymptotic constant factor difference or finite size effects between matrix inversion and matrix multiplication. We can find the threshold where these costs are comparable, that is, the minimum number of qubits for which the classical cost of VQS times the prefactor is smaller than the cost of the purely classical approach. We plot this in figure~\ref{fig:threshold} as a function of the prefactor $p$ within a certain range. We observe that the growth of this threshold slows down rapidly as $p$ grows. 

This speaks in favor of the applicability of VQS as an alternative to purely classical methods with exploding computational cost in system size. We can therefore identify a system-size-dependent corridor in the number of qubits for the applicability of VQS: large enough to use fewer classical computational resources than the purely classical approach, and small enough as to get better performance than Trotter. Looking at both figure~\ref{fig:extrapolation_fit} and figure~\ref{fig:threshold}, the trend suggests that for a long-enough simulated time, this advantage corridor is likely to be non-empty, even being pessimistic about the value of the relative prefactor linking the two computational costs.

\section{Conclusions and outlook}

We performed a computational analysis of VQS and compared it with a second-order Trotter formula, a well-established quantum algorithm for time evolution, in order to verify the alleged shallowness of circuits involved in VQS empirically. This comparative study was motivated by a trend in the literature to look for quantum advantage scenarios~\cite{future_ibm, Childs_useful}, especially for such a promising field as quantum dynamics simulation, due to the natural mapping of these problems to quantum computers. The results suggest that while it might not be advantageous using the VQS algorithm for short-time simulations, because the depth requirement for VQS is not substantially lower than Trotter's, for longer times VQS might have an advantage. This is a consequence of the fact that the minimum depth required to reach a desired fidelity appears to grow slower in simulated time for VQS, as seen in figure~\ref{fig:tfs_comparison}. In the current (pre-fault-tolerant) era of quantum computing, where noise plays a deciding role in the viability of algorithms, such reductions in depth could be the difference between a high-quality and a low-quality simulation. For many physical systems, such as those reaching a steady-state or presenting chaotic behavior after a certain time, only a long-time simulation could in general capture the main dynamical properties.

However, we reiterate that these results are only empirical, and a full theoretical treatment of VQS is still lacking. The linear system of equations in the VQS algorithm is susceptible to numerical instabilities due to the ill-conditioning of the matrix $\mathcal{A}$, which is problem- and ansatz-dependent and restricts the applicability of VQS in a rather hard-to-predict way.

Possible future directions include therefore a more detailed performance study taking into account not only the circuit depth, which is the dominant limitation on quantum algorithm implementation on today's devices, but also other second-order contributions to the overall computational cost, such as the number of circuits that need to be computed in VQS, which is directly related to the classical cost of solving the ill-conditioned linear system of equations. The noise resilience of this algorithm is also still to be demonstrated, most probably determined by the chosen overlap-evaluation technique. The parameter update rule for longer and longer simulated times could end up being dominated by the noisy measurements by which previous updates where made. However, variational methods might have the potential to better neutralize control errors, like a constant over- or under-rotation~\cite{vqa_noise1,vqa_noise2}. All these would shed more light into the applicability of VQS as an alternative to Trotter for long-time quantum simulation.

\section*{References}
\bibliographystyle{iopart-num}
\bibliography{library} 

\providecommand{\newblock}{}
\begin{thebibliography}{10}
\expandafter\ifx\csname url\endcsname\relax
  \def\url#1{{\tt #1}}\fi
\expandafter\ifx\csname urlprefix\endcsname\relax\def\urlprefix{URL }\fi
\providecommand{\eprint}[2][]{\url{#2}}

\bibitem{preskill}
Preskill J 2018 {\em {Quantum}\/} {\bf 2} 79 ISSN 2521-327X \urlprefix\url{https://doi.org/10.22331/q-2018-08-06-79}

\bibitem{chemist}
Georgescu I~M, Ashhab S and Nori F 2014 {\em Rev. Mod. Phys.\/} {\bf 86}(1) 153--185 \urlprefix\url{https://link.aps.org/doi/10.1103/RevModPhys.86.153}

\bibitem{classical}
Fornberg B 1996 {\em A Practical Guide to Pseudospectral Methods\/} Cambridge Monographs on Applied and Computational Mathematics (Cambridge University Press)

\bibitem{univ_lloyd}
Lloyd S 1996 {\em Science\/} {\bf 273} 1073--1078 (\textit{Preprint} \eprint{https://www.science.org/doi/pdf/10.1126/science.273.5278.1073}) \urlprefix\url{https://www.science.org/doi/abs/10.1126/science.273.5278.1073}

\bibitem{future_ibm}
Bravyi S, Dial O, Gambetta J~M, Gil D and Nazario Z 2022 {\em Journal of Applied Physics\/} {\bf 132} 160902 ISSN 0021-8979 (\textit{Preprint} \eprint{https://pubs.aip.org/aip/jap/article-pdf/doi/10.1063/5.0082975/16515734/160902\_1\_online.pdf}) \urlprefix\url{https://doi.org/10.1063/5.0082975}

\bibitem{trotter_error}
Childs A~M, Su Y, Tran M~C, Wiebe N and Zhu S 2021 {\em Phys. Rev. X\/} {\bf 11}(1) 011020 \urlprefix\url{https://link.aps.org/doi/10.1103/PhysRevX.11.011020}

\bibitem{berry_exp}
Berry D~W, Childs A~M, Cleve R, Kothari R and Somma R~D 2014 Exponential improvement in precision for simulating sparse hamiltonians {\em Proceedings of the Forty-Sixth Annual ACM Symposium on Theory of Computing\/} STOC '14 (New York, NY, USA: Association for Computing Machinery) p 283–292 ISBN 9781450327107 \urlprefix\url{https://doi.org/10.1145/2591796.2591854}

\bibitem{asympt}
Haah J, Hastings M~B, Kothari R and Low G~H 2023 {\em SIAM Journal on Computing\/} {\bf 52} FOCS18--250--FOCS18--284 (\textit{Preprint} \eprint{https://doi.org/10.1137/18M1231511}) \urlprefix\url{https://doi.org/10.1137/18M1231511}

\bibitem{variational}
Cerezo M, Arrasmith A, Babbush R, Benjamin S~C, Endo S, Fujii K, McClean J~R, Mitarai K, Yuan X, Cincio L and Coles P~J 2021 {\em Nature Reviews Physics\/} {\bf 3} 625–644 ISSN 2522-5820 \urlprefix\url{http://dx.doi.org/10.1038/s42254-021-00348-9}

\bibitem{vqs1}
Li Y and Benjamin S~C 2017 {\em Phys. Rev. X\/} {\bf 7}(2) 021050 \urlprefix\url{https://link.aps.org/doi/10.1103/PhysRevX.7.021050}

\bibitem{vqs2}
McArdle S, Jones T, Endo S, Li Y, Benjamin S~C and Yuan X 2019 {\em npj Quantum Information\/} {\bf 5} ISSN 2056-6387 \urlprefix\url{http://dx.doi.org/10.1038/s41534-019-0187-2}

\bibitem{vqs4}
Endo S, Sun J, Li Y, Benjamin S~C and Yuan X 2020 {\em Phys. Rev. Lett.\/} {\bf 125}(1) 010501 \urlprefix\url{https://link.aps.org/doi/10.1103/PhysRevLett.125.010501}

\bibitem{adaptvqe}
Grimsley H~R, Economou S~E, Barnes E and Mayhall N~J 2019 {\em Nature Communications\/} {\bf 10} ISSN 2041-1723 \urlprefix\url{http://dx.doi.org/10.1038/s41467-019-10988-2}

\bibitem{adaptvarrte}
Yao Y~X, Gomes N, Zhang F, Wang C~Z, Ho K~M, Iadecola T and Orth P~P 2021 {\em PRX Quantum\/} {\bf 2} ISSN 2691-3399 \urlprefix\url{http://dx.doi.org/10.1103/PRXQuantum.2.030307}

\bibitem{adaptvarqite}
Gomes N, Mukherjee A, Zhang F, Iadecola T, Wang C, Ho K, Orth P~P and Yao Y 2021 {\em Advanced Quantum Technologies\/} {\bf 4} ISSN 2511-9044 \urlprefix\url{http://dx.doi.org/10.1002/qute.202100114}

\bibitem{vqs3}
Yuan X, Endo S, Zhao Q and Li Yiand~Benjamin S~C 2019 {\em Quantum\/} {\bf 3} 191 ISSN 2521-327X \urlprefix\url{http://dx.doi.org/10.22331/q-2019-10-07-191}

\bibitem{error_bounds}
Zoufal C, Sutter D and Woerner S 2023 {\em Physical Review Applied\/} {\bf 20} ISSN 2331-7019 \urlprefix\url{http://dx.doi.org/10.1103/PhysRevApplied.20.044059}

\bibitem{trotter1}
Hatano N and Suzuki M 2005 {\em Finding Exponential Product Formulas of Higher Orders\/} (Springer Berlin Heidelberg) p 37–68 ISBN 9783540315155 \urlprefix\url{http://dx.doi.org/10.1007/11526216_2}

\bibitem{trotter2}
Magnus W 1954 {\em Communications on Pure and Applied Mathematics\/} {\bf 7} 649--673 (\textit{Preprint} \eprint{https://onlinelibrary.wiley.com/doi/pdf/10.1002/cpa.3160070404}) \urlprefix\url{https://onlinelibrary.wiley.com/doi/abs/10.1002/cpa.3160070404}

\bibitem{Childs_useful}
Childs A~M, Maslov D, Nam Y, Ross N~J and Su Y 2018 {\em Proceedings of the National Academy of Sciences\/} {\bf 115} 9456–9461 ISSN 1091-6490 \urlprefix\url{http://dx.doi.org/10.1073/pnas.1801723115}

\bibitem{qgeom}
Haug T, Bharti K and Kim M 2021 {\em PRX Quantum\/} {\bf 2}(4) 040309 \urlprefix\url{https://link.aps.org/doi/10.1103/PRXQuantum.2.040309}

\bibitem{tikhonov}
Tikhonov A~N 1977 {\em Solutions of ill-posed problems\/} Scripta series in mathematics (Washington: V. H. Winston I\& Sons) ISBN 0-470-99124-0

\bibitem{numpy}
Harris C~R, Millman K~J, van~der Walt S~J, Gommers R, Virtanen P, Cournapeau D, Wieser E, Taylor J, Berg S, Smith N~J, Kern R, Picus M, Hoyer S, van Kerkwijk M~H, Brett M, Haldane A, del R{\'{i}}o J~F, Wiebe M, Peterson P, G{\'{e}}rard-Marchant P, Sheppard K, Reddy T, Weckesser W, Abbasi H, Gohlke C and Oliphant T~E 2020 {\em Nature\/} {\bf 585} 357--362 \urlprefix\url{https://doi.org/10.1038/s41586-020-2649-2}

\bibitem{rk45}
Fehlberg E 1964 {\em ZAMM - Journal of Applied Mathematics and Mechanics / Zeitschrift für Angewandte Mathematik und Mechanik\/} {\bf 44} T17--T29 (\textit{Preprint} \eprint{https://onlinelibrary.wiley.com/doi/pdf/10.1002/zamm.19640441310}) \urlprefix\url{https://onlinelibrary.wiley.com/doi/abs/10.1002/zamm.19640441310}

\bibitem{scipy}
Virtanen P, Gommers R, Oliphant T~E, Haberland M, Reddy T, Cournapeau D, Burovski E, Peterson P, Weckesser W, Bright J, {van der Walt} S~J, Brett M, Wilson J, Millman K~J, Mayorov N, Nelson A~R~J, Jones E, Kern R, Larson E, Carey C~J, Polat {\.I}, Feng Y, Moore E~W, {VanderPlas} J, Laxalde D, Perktold J, Cimrman R, Henriksen I, Quintero E~A, Harris C~R, Archibald A~M, Ribeiro A~H, Pedregosa F, {van Mulbregt} P and {SciPy 10 Contributors} 2020 {\em Nature Methods\/} {\bf 17} 261--272

\bibitem{hva1}
Wecker D, Hastings M~B and Troyer M 2015 {\em Physical Review A\/} {\bf 92} ISSN 1094-1622 \urlprefix\url{http://dx.doi.org/10.1103/PhysRevA.92.042303}

\bibitem{hva2}
Wiersema R, Zhou C, de~Sereville Y, Carrasquilla J~F, Kim Y~B and Yuen H 2020 {\em PRX Quantum\/} {\bf 1} ISSN 2691-3399 \urlprefix\url{http://dx.doi.org/10.1103/PRXQuantum.1.020319}

\bibitem{hva3}
Park C~Y and Killoran N 2024 {\em Quantum\/} {\bf 8} 1239 ISSN 2521-327X \urlprefix\url{http://dx.doi.org/10.22331/q-2024-02-01-1239}

\bibitem{qulacs}
Suzuki Y, Kawase Y, Masumura Y, Hiraga Y, Nakadai M, Chen J, Nakanishi K~M, Mitarai K, Imai R, Tamiya S, Yamamoto T, Yan T, Kawakubo T, Nakagawa Y~O, Ibe Y, Zhang Y, Yamashita H, Yoshimura H, Hayashi A and Fujii K 2021 {\em Quantum\/} {\bf 5} 559 ISSN 2521-327X \urlprefix\url{http://dx.doi.org/10.22331/q-2021-10-06-559}

\bibitem{qutip1}
Johansson J, Nation P and Nori F 2012 {\em Computer Physics Communications\/} {\bf 183} 1760--1772 ISSN 0010-4655 \urlprefix\url{https://www.sciencedirect.com/science/article/pii/S0010465512000835}

\bibitem{qutip2}
Johansson J, Nation P and Nori F 2013 {\em Computer Physics Communications\/} {\bf 184} 1234--1240 ISSN 0010-4655 \urlprefix\url{https://www.sciencedirect.com/science/article/pii/S0010465512003955}

\bibitem{vqa_noise1}
Liang Z, Cheng J, Ren H, Wang H, Hua F, Song Z, Ding Y, Chong F, Han S, Qian X and Shi Y 2024 Napa: Intermediate-level variational native-pulse ansatz for variational quantum algorithms (\textit{Preprint} \eprint{2208.01215})

\bibitem{vqa_noise2}
Skolik A, Mangini S, B{\"a}ck T, Macchiavello C and Dunjko V 2023 {\em EPJ Quantum Technology\/} {\bf 10} 8 ISSN 2196-0763 \urlprefix\url{https://doi.org/10.1140/epjqt/s40507-023-00166-1}

\end{thebibliography}

\end{document}